\documentclass[12pt]{article}

\usepackage{natbib}
\usepackage{url}
\usepackage{graphicx,epstopdf}
\usepackage{enumitem}
\usepackage{amsmath,amsthm}
\include{ThermBib.bib}

\newcommand{\dbar}{d\mkern-6mu\mathchar'26}
\newcommand{\ex}[1]{\langle #1 \rangle}
\newcommand{\mb}{\mathbf}
\newcommand{\mc}{\mathcal}

\newtheorem{proposition}{Proposition}
\newtheorem{corollary}{Corollary}[proposition]
\newtheorem{lemma}{Lemma}
\newcommand{\bibsuffix}[1]{}

\title{Shakin' All Over: Proving  Landauer's Principle without neglect of fluctuations}
\author{Wayne C. Myrvold \\ Department of Philosophy \\ The University of Western Ontario \\ wmyrvold@uwo.ca}

\begin{document}
\maketitle
\abstract{Landauer's principle is, roughly,  the principle that there is an entropic cost associated with implementation of logically irreversible operations. Though widely accepted in the literature on the thermodynamics of computation, it has been the subject of considerable dispute in the philosophical literature. Both the cogency of proofs of the principle and its relevance, should it be true, have been questioned.  In particular, it has been argued that microscale fluctuations  entail dissipation that always greatly exceeds the Landauer bound.  In this article Landauer's principle is treated within statistical mechanics, and a proof is given that neither relies on neglect of fluctuations nor assumes the availability of thermodynamically reversible processes.  In addition, it is argued that microscale fluctuations are no obstacle  to approximating  thermodynamic reversibility as closely as one would like.}

\section{Introduction} The statement that has come to be known as \emph{Landauer's Principle} is, roughly, that there is an entropic cost associated with implementation of  logically irreversible operations, that is, operations whose input states cannot be recovered from their output states.  It is widely accepted in the literature on the thermodynamics of computation; see \citet{LeffRex2} for a sampling of the relevant literature and an extensive bibliography.  Nonetheless, it has been the subject of considerable controversy in the philosophical literature \citep{ExorcistII,LotusEaters,LPSG2007,LPS2008,NortonWaiting,LRDefended,NortonReply,NortonNoGo,NortonShook,LRCircles,Ladyman2018,Norton2018}.

Ladyman, Presnell,  Short, and Groisman (\citeyear{LPSG2007}), hereinafter referred to as LPSG, presented  a proof of Landauer's principle. The proof, like any proof, rests on assumptions.  The operative assumptions of the proof are that a probabilistic version of the second law of thermodynamics holds, and that certain processes can be performed reversibly. These processes include, crucially, expansion of a single-molecule  gas.  \citet{NortonWaiting,NortonNoGo,NortonShook}  has argued that inevitable fluctuations at the molecular level invalidate  the assumption of even approximate thermodynamic reversibility of processes at the microscale,   and  that any process involves dissipation in excess of the bounds required by  Landauer's principle, rendering the principle moot.   This is regarded by Norton as a `no-go' result, invalidating the basic framework within which most of the work on thermodynamics of computation has been carried out.

\citet{LRCircles}  addressed the  purported no-go result, arguing that the conclusion has not been established.  They acknowledged, however, a concern about the  assumption, ubiquitous in the literature on thermodynamics of computation, of molecular-scale processes carried out with negligible dissipation.

In this article, the subject of Landauer's principle is addressed from the point of view of statistical mechanics.  It is shown that the relevant version of the second law of thermodynamics is provable within statistical mechanics, in two versions,  classical and quantum. It  is therefore not required as an independent assumption.  A derivation within statistical mechanics of the Landauer principle is given, that neither relies on neglect of fluctuations nor  assumes the availability of thermodynamically reversible processes.

As Norton has rightly emphasized, a theorem of this sort is moot if the processes involved depart sufficiently far from thermodynamic reversibility. This is explicit in the theorem we  prove. Unless there are  processes available that  approximate reversibility sufficiently closely, the theorem places no bounds on \emph{extra} dissipation associated with logical irreversibility. For that reason, I will argue that, given the notion of thermodynamic reversibility relevant to the context at hand,  fluctuations, even ones that are large on the scale at which the processes are taking place, pose no threat to the assumption that processes can take place that approximate thermodynamic reversibility as closely as we would like.

\section{The set-up}\label{setup}  As is usual in thermodynamics, the thermodynamic state of a system $A$ is defined with respect to some set of manipulable variables $\boldsymbol{\lambda} = \{\lambda_1, \lambda_2, \ldots, \lambda_n \}$, which may represent, for example, the positions of the walls of a container the system is constrained to be in, or the value of applied fields. We thus consider a family of Hamiltonians $\{H_{\boldsymbol{\lambda}}\}$.  The variables $\boldsymbol{\lambda}$  are treated as exogenous, meaning that we do not include in our physical description the systems that are the sources of these applied fields, and we do not consider the influence of the system $A$ on those systems.  They may also be freely specified, independently of the state of the system.  We consider some set $\mc{M}$ of manipulations of the system, where each manipulation consists of some specification of  $\boldsymbol{\lambda}(t)$ through some interval $t_0 \leq t \leq t_1$. In addition, we may assume that there are available one or more heat reservoirs $\{B_i\}$ at temperatures $T_i$, with which the system can exchange heat.  The system $A$ may be coupled and decoupled from these heat baths during the course of its evolution. That is, the interaction terms in the total Hamiltonian consisting of the system $A$ and the reservoirs  $\{B_i\}$ are also treated as manipulable variables.

Since the time of \citet{MaxTOH,TaitII} it has been recognized  that the kinetic theory of heat entails that the second law of thermodynamics, as originally formulated, cannot hold strictly. When compressing a gas with a piston, we might   find on some occasion that, due to a fluctuation in the force exerted by the gas on the piston,  less work is needed to compress the gas than one would expect on average, and so in a given cycle of a heat engine might obtain more work than is allowed by the second law from a given quantity of heat extracted.  By the same token we might obtain less work than  expected.  We do not, however, expect that we will be able to \emph{consistently and reliably} violate the Carnot limit on efficiency of a heat engine.  The original version of the second law should be replaced by a probabilistic one. The second law will then be, to employ Szilard's vivid analogy, like a theorem about the impossibility of a gambling system intended to beat the odds set by a casino.
\begin{quote}
Consider somebody playing a thermodynamical gamble with the help of cyclic processes and with the intention of decreasing the entropy of the heat reservoirs. Nature will deal with him like a well established casino, in which it is possible to make an occasional win but for which no system exists ensuring the gambler a profit (\citealt[p. 73]{Szilard1925e}, from \citealt[p. 757]{Szilard1925}).
\end{quote}
On a macroscopic scale, we expect  fluctuations to be negligible, but, as Norton has  emphasized,  at the microscale on which in-principle limitations on the thermal cost of computation are  investigated, they are far from negligible.  Accordingly, we will invoke probabilistic considerations, and treat of the evolution of probability distributions over the state of a system subjected to various manipulations.  When considering the amount of work needed to perform an operation, or the amount of heat exchanged in the course of the evolution of the system, we will consider \emph{expectation values} of work and heat exchanges, calculated with respect to those probability distributions.

We assume it makes sense to associate a probability distribution with a preparation procedure, and to compute on its basis probabilities for outcomes of subsequent manipulations. We need not enquire into the status of these probabilities, so long as they serve this purpose.

Since the late nineteenth century it has been common to think of probability statements as involving veiled reference to relative frequencies in an actual or hypothetical sequence of events, or in an ensemble of similarly prepared systems.  There is no commitment here to such frequentism about probabilities; probability considerations may be applied to single events. There is, however, a link between probabilities and mean outcomes in a long sequence of trials, afforded by the weak law of large numbers. Suppose that we are able to conduct multiple runs of a procedure, in such a way that the probabilities of the outcomes are the same on each trial, and  the outcomes of any trial are probabilistically independent of the outcomes of all the others. Then, if we take any outcome variable, and compute its mean value across the results of a long sequence of trials, with high probability this mean value will be close to its expectation value on a single trial.  We can make the probability of any degree of approximation  as high as we like by increasing the number of trials. Though the expectation values to be invoked are not \emph{defined} in terms of  mean values in a long sequence of trials, they have implications for such  mean values.  If we could construct a heat engine  such that the expectation value of work extracted on each run exceeded the Carnot bound, we could, by running sufficiently many cycles, make the probability of a net violation of the Carnot bound as close to unity as we like.

We will treat of ``states'' $a = ( \rho_a, H_a )$,  consisting of a probability distribution over the phase space of the system $A$ (or, in the quantum context, a density operator  on the system's Hilbert space), represented by a density $\rho_a$, sand a Hamiltonian, which, as already noted, may depend on  exogenous, manipulable variables. We  consider the effects on those states of manipulations in some class $\mc{M}$.

As is usual in statistical mechanics, the distributions associated  with the heat reservoirs $B_i$ will be canonical distributions, uncorrelated with the system $A$ (see \citealt{MaroneyEntropy} for discussion of the justification for this use of canonical distributions). In the classical context, a canonical distribution is a  distribution that has density, with respect to Liouville measure,
\begin{equation}
\rho_\beta = Z^{-1} e^{-\beta H},
\end{equation}
where $\beta$ is the inverse temperature $1/kT$, and $Z$ is the normalization constant required to make the integral of this density over all phase space unity.  This depends both on the Hamiltonian $H$ and on $\beta$, and is called the \emph{partition function}.  In the quantum context, a canonical state is represented by density operator
\begin{equation}
\hat{\rho} _\beta = Z^{-1} e^{-\beta \hat{H}},
\end{equation}
where, again, $Z$ is the constant required to normalize the state.

As the reservoirs interact with $A$, correlations will be built up, but we will assume  that the reservoirs are big enough and noisy enough that these are, as far as subsequent interactions with $A$  are concerned, effectively effaced, meaning that the effect of the reservoirs on $A$ is \emph{as if} they are uncorrelated. This means, not that the probability distribution over the full state of $A$ and $B_i$ is a product distribution, but that the dynamical variables of $B_i$ and $A$ relevant to interactions their interactions with $A$ are effectively independent.

The manipulations of a system $A$ we will be considering will be ones of the following form.
\begin{itemize}
\item At time $t_0$, the system has some probability distribution $\rho_a$, and the Hamiltonian of the system $A$ is $H_a$.
\item At time $t_0$, the heat reservoirs $B_i$  have canonical distributions at temperatures $T_i$, uncorrelated with $A$, and are not interacting with $A$.
\item During the time interval $[t_0, t_1]$, the composite system consisting of $A$ and the reservoirs $\{B_i\}$ undergoes Hamiltonian evolution, governed by a time-dependent Hamiltonian $H(t)$,  which may include successive couplings between $A$ and the heat reservoirs $\{B_i\}$.
\item The internal Hamiltonians of the reservoirs $\{B_i\}$ do not change.
\item At time $t_1$, the Hamiltonian of the system $A$ is $H_b$, and, as a result of Hamiltonian evolution of the composite system, the marginal probability distribution of $A$ is $\rho_b$.
\end{itemize}
This is a manipulation that takes a state $a = ( \rho_a, H_a )$ to state $b = ( \rho_b, H_b )$.

It should be noted that we are \emph{not} considering manipulations that consist of a measurement performed on the system $A$ followed by a manipulation of the exogenous variables whose choice depends on the outcome of the measurement. Controlled operations are allowed, but the control mechanism must be internalized, that is, included in the system under study.  The system $A$ could consist of two parts $A_1$ and $A_2$, which interact in such a way that the state of $A_1$  affects what happens to $A_2$, which subsequently affects what happens to  $A_1$. But all of this must be encoded in the Hamiltonian $H(t)$, which may be time-varying but which undergoes a preprogrammed evolution that is \emph{not} dependent on the state of the system $A$.  Otherwise, there may be dissipation associated with the operation of the control mechanism that gets left out of the analysis.

We will count energy exchanges with the reservoirs $B_i$ as heat (to be counted as positive if $A$ gains energy from $B$, negative if $A$ loses energy), and energy changes to $A$ due to changes in the external potentials as work (again, counted as positive if $A$ gains energy, negative if it loses energy).

Dropping the assumption of the availability of reversible processes requires revision of the familiar framework of thermodynamics, as it means dropping the assumption of the availability of an entropy function.  In its place we will define quantities $S_\mc{M}(a \rightarrow b)$, defined relative to a class of available manipulations $\mc{M}$, to be thought of as analogs, in the current context, of entropy differences between states $a$ and $b$. These will be representable as differences in the values of some state function only in the limiting case in which all states can be connected reversibly.

For any manipulation $M$, that takes a state $a$ to a state $b$, we can define $\ex{Q_i(a \rightarrow b)}_M$ as the expectation value of the heat obtained by $A$ from reservoir $B_i$.  We can use these to define,
\begin{equation}
\sigma_M(a\rightarrow b)  = \sum_i \frac{\ex{Q_i(a \rightarrow b)}_M}{T_i}.
\end{equation}
Let $\mc{M}(a\rightarrow b)$ be the set of manipulations in $\mc{M}$ that take $a$ to $b$, and define, as analogs of entropies (which we will henceforth just call ``entropies''),
\begin{equation}
S_\mc{M}(a \rightarrow b) =  \mbox{l.u.b.}\{ \sigma_M(a\rightarrow b) \: | \, M \in \mc{M}(a \rightarrow b) \}.
\end{equation}
Via the obvious extension of this definition we also define quantities such as  $S_\mc{M}(a \rightarrow b \rightarrow c)$ for processes with any number of intermediate steps.  It is assumed that manipulations can be composed, that is, that any manipulation that takes $a$ to $b$ can be followed by one that takes $b$ to $c$ to form a manipulation that takes $a$ to $b$ and then to $c$.  It follows from this composition assumption and the definition of the entropies that
\begin{equation}
S_\mc{M}(a \rightarrow b \rightarrow c) = S_\mc{M}(a \rightarrow b) + S_\mc{M}(b \rightarrow c),
\end{equation}
and similarly for processes consisting of longer chains of intermediates states.

One version of the second law of thermodynamics says that, for any cyclic process, the sum of $Q_i/T_i$ over all heat exchanges cannot be positive.  Since we're working in the context of statistical mechanics, and we do not want to ignore fluctuations, the appropriate revision of the second law involves expectation values of heat exchanges.  A cyclic process will be one that restores the marginal probability distribution of the system $A$ to the one it started out with.  The revised second law that we  will prove in the next section states that, for any cyclic process, the sum of $\ex{Q_i}/T_i$  over all heat exchanges cannot be positive. In the notation we have introduced, this is:
\begin{quote} \textbf{The Statistical Second Law}. For any state $a$,
\[
S_\mc{M}(a \rightarrow a) \leq 0.
\]
\end{quote}
It follows from this that
\begin{equation}
S_\mc{M}(a \rightarrow b \rightarrow a) = S_\mc{M}(a \rightarrow b) + S_\mc{M}(b \rightarrow a) \leq 0,
\end{equation}
and similarly for processes involving longer chains of intermediate states.

In any process $M$ that takes a state $a$ to a state $b$, some of the work done, or heat discarded into a reservoir, may be recovered by some process that takes $b$ back to $a$.  If the process can be reversed with the signs of all $\ex{Q_i}$ reversed, then full recovery is possible.  If full recovery is not possible, and cannot even be approached arbitrarily closely, we will say that the process is \emph{dissipatory}.  A manipulation $M'$ that takes $b$ to $a$ and recovers work done and heat discarded would be one such that
\begin{equation}
\sigma_M(a \rightarrow b) + \sigma_{M'}(b \rightarrow a) = 0.
\end{equation}
There might be a limit to how closely this can be approached. Define the dissipation associated with the process of $M$ taking $a$ to $b$ as the distance between this limit and perfect recovery.
\begin{align}\label{dissdef}
\nonumber \delta_M(a \rightarrow b) &= \mbox{g.l.b.}\{ -(\sigma_M(a \rightarrow b) + \sigma_{M'}(b \rightarrow a)) \: | \, M' \in \mc{M}(b \rightarrow a) \}
\\ &= -S_\mc{M}(b \rightarrow a) - \sigma_M(a \rightarrow b).
\end{align}
It follows from the statistical second law that this is non-negative.

If there is no limit to how much the dissipation associated with processes that connect $a$ to $b$ can be diminished,
\begin{equation}\label{reversible}
S_\mc{M}(a \rightarrow b \rightarrow a) = 0.
\end{equation}
When this  holds, it is traditional to say that $a$ and $b$ can be connected reversibly, and to imagine a fictitious process that can proceed in either direction, reversing the signs of all heat exchanges.  There is no harm in doing so, as long as this is not taken too literally.\footnote{As \citet{ImpossibleProcess} has argued, taking talk of irreversible processes too literally can lead to contradictions.}  Following convention, we will say, for any $a$, $b$ for which (\ref{reversible}) is satisfied, that $a$ and $b$ can be connected reversibly.  When this locution is used, bear in mind that it is shorthand for (\ref{reversible}), and does not presume the existence of an actual reversible process.

From the statistical second law it follows that, if all states can be connected reversibly---that is, if, for all $a, b$, $S_\mc{M}(a \rightarrow b \rightarrow a) = 0$---then there exists a state function $S_\mc{M}$, defined up to an additive constant, such that
\begin{equation}
S_\mc{M}(a \rightarrow b) = S_\mc{M}(b)-  S_\mc{M}(a).
\end{equation}
This is the familiar entropy function. The reason we have been expressing things in an unfamiliar way is that we \emph{don't} want to assume reversibility as a general rule.

Any manipulation that takes $a$ to $b$ must have dissipation of at least  $-S_\mc{M}(a \rightarrow b \rightarrow a)$.   Define the \emph{inefficiency} associated with a manipulation that takes $a$ to $b$  as the amount by which its dissipation exceeds this minimal value.
\begin{align}
\nonumber \eta_M(a \rightarrow  b) &= \delta_M(a \rightarrow b) - (-S_\mc{M}(a \rightarrow b \rightarrow a))
\\ &= S_\mc{M}(a \rightarrow b) - \sigma_M(a \rightarrow b).
\end{align}
If $a$ and $b$ can be connected reversibly, the distinction between dissipation and inefficiency vanishes, and the inefficiency is equal to the dissipation.

We are now in a position to state the version of Landauer's principle that we will be proving.  Consider a logical operation $L$ that is not logically reversible, meaning that the input is not recoverable from the output. This means that there are two or more inputs $\{ \alpha_i \}$ that are mapped by $L$ to the same output $\beta$.  In a  device that implements the logical operation $L$, the inputs will be represented by statistical mechanical states $\{a_i\}$, and the output by a statistical mechanical state $b$.  Distinct inputs are to be represented by distinguishable states, which, in the classical context, means probability distributions with non-overlapping support, and in the quantum-mechanical context, by orthogonal density operators.  An implementation of $L$ is a manipulation $M_L$ that maps each member of the set  $\{a_i\}$ into $b$.

The question to be asked is: can the manipulation $M_L$ do this without incurring any inefficiency?  That is, can we have $\eta_{M_L}(a_i \rightarrow b)$ equal to zero, for every $a_i$?  Failing that, can we, by appropriate choice of manipulation, make every element of the set  $\{\eta_{M_L}(a_i \rightarrow b)\}$ arbitrarily small?

In the next section we will prove the following.

\begin{figure}[ht]
\centering
\includegraphics[width=0.4\textwidth]{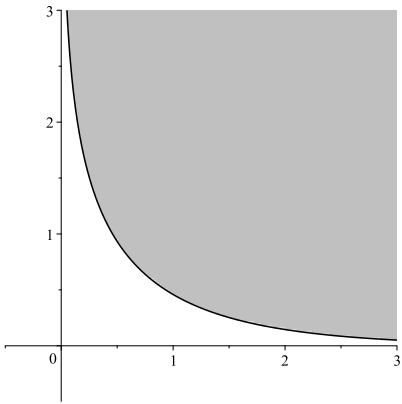}
\caption{Values of $(\delta_1, \delta_2)$ permitted by Landauer's principle.}\label{Lconstraint}
\end{figure}

\begin{quote}\textbf{Landauer bound on dissipations.}  If manipulation $M$ takes each of a distinguishable set $\{a_i, i = 1, \ldots, n\}$ of states to the same state $b$, then
\[
\sum_{i=1}^n \: e^{-\delta_M(a_i \rightarrow b)/k}  \leq 1.
\]
\end{quote}
 This  entails that every member of the set $\{\delta_{M}(a_i \rightarrow b)\}$ is greater than zero.  It also entails a formulation that is often presented as a gloss of Landauer's principle, that the mean of the set is not  smaller than $k \log{n}$.\footnote{In this article, all logarithms are natural logarithms, that is, logarithms to the base $e$.}
\begin{equation}
\frac{1}{n} \sum_{i=1}^n \delta_{M}(a_i \rightarrow b) \geq k \log n.
\end{equation}
That is, there an \emph{average} dissipation, taken over members of the set $\{a_i\}$, of at least $k \log n$.\footnote{See Appendix for proof that this is entailed by our formulation of the Landauer principle.}  We might be able to reduce the dissipation  associated with any particular member of the set as much as we like, but we cannot simultaneously make all of them arbitrarily small. For the case of $n = 2$, the most commonly discussed case, the constraint is graphed in Figure \ref{Lconstraint}. The shaded region is the set of permitted pairs $(\delta_1, \delta_2) = (\delta_{M}(a_1 \rightarrow b)/k, \delta_{M}(a_2 \rightarrow b)/k)$.

If, as is usually assumed in these discussions, the states $\{a_i\}$ can be connected reversibly to $b$, then any dissipation is inefficiency, and bounds on dissipations are bounds on inefficiencies.  If reversibility is not assumed, there may be unavoidable levels of dissipation associated with some state transitions; if this is the case, not every dissipation represents an inefficiency.  We can re-state the Landauer principle in terms of inefficiencies.

\begin{quote}\textbf{Landauer bound on inefficiencies.} If manipulation $M$ takes each of a distinguishable set $\{a_i, i = 1, \ldots, n\}$ of states to the same state $b$, then
\[
\sum_{i=1}^n \: e^{-\left(\eta_i - S_\mc{M}(a_i \rightarrow b \rightarrow a_i)\right)/k}  \leq 1,
\]
where $\eta_i$ is the inefficiency $\eta_{M}(a_i \rightarrow b)$.
\end{quote}
If we have reversibility, then this entails that all of the inefficiencies $\eta_{M}(a_i \rightarrow b)$  must be positive, and that they cannot all be made arbitrarily small in the same process.  Far enough from reversibly, it places no constraint on inefficiencies at all.  The condition for the Landauer bound to place a constraint on inefficiencies is,
\begin{equation}\label{shook1}
\sum_{i =1}^n e^{S_\mc{M}(a_i\rightarrow b \rightarrow a_i)/k} >  1.
\end{equation}
A necessary condition for (\ref{shook1})  to be satisfied, and thus for the Landauer principle to have teeth, is the condition that, for some $a_i$,
\begin{equation}\label{shook2}
S_\mc{M}(a_i \rightarrow b \rightarrow a_i) > -k \log n.
\end{equation}
If \citet{NortonWaiting,NortonNoGo,NortonShook} is right about the minimum dissipation required for  carrying out processes at the molecular level, then  (\ref{shook2}) is \emph{not} satisfiable; because of fluctuations at the molecular level, any process departs from reversibility by an amount that far exceeds the Landauer bound.  In section \ref{reverse} it will be argued  that this is not correct, and the Landauer principle does have teeth.

The Landauer bound  we have stated  involves a distinguishable set of states.  Distinguishability, like reversibility, is something that we should not expect to hold perfectly; in actual implementations we will at best approximate perfect distinguishability.  For this reason, the theorem that we will  prove in the next section will not require perfect distinguishability, and will entail the version of the Landauer bound  we have stated in this section as a special case.

\section{Proving the Second Law, and Landauer's Principle}\label{proving}
The theorems we will be concerned with come in two  versions, classical and quantum, each proven in pretty much the same way. To avoid saying everything twice, we adopt a systematically ambiguous notation, and state each theorem in such a way that it can be read either as a theorem of classical statistical mechanics, or as a theorem of quantum statistical mechanics.

In what follows, $\rho$ will be used either for a density function, with respect to Liouville measure, on a  classical phase space, or, in the quantum context,  a density operator on a Hilbert space.  $S[\rho]$ is  the Gibbs entropy  (classical), or  the von Neumann entropy (quantum).
\begin{equation}
S[\rho] = -k \, \ex{\log \rho}_\rho.
\end{equation}
We also define the relative entropy of two distributions.
\begin{equation}
S[ \rho \, \|  \, \sigma] = -k\left( \, \ex{\log \sigma}_\rho - \ex{\log \rho}_\rho   \right).
\end{equation}
$S[ \rho \, \|  \, \sigma]$ is one way to  measure how much the distribution represented by $\sigma$ departs from that represented by $\rho$. It is equal to zero for $\sigma = \rho$, and is positive for any other $\sigma$.

Suppose $\bar{a}$ is a probabilistic mixture of states $\{ a_i\}$.
\begin{equation}
\rho_{\bar{a}} = \sum_{i = 1}^n p_i \,\rho_{a_i},
\end{equation}
where $\{p_i\}$ are positive numbers that add up to one. Then the Gibbs/von Neumann entropy of $\bar{a}$ is related to that of the $a_i$'s via,
\begin{equation}\label{mix}
S[\rho_{\bar{a}}] = \sum_{i = a}^n p_i \, S[\rho_{a_i}] +   \sum_{i = a}^n p_i \, S[ \rho_{a_i} \, \| \, \rho_{\bar{a}} ].
\end{equation}
If the states $\{ a_i\}$ are distinguishable, then  $S[ \rho_{a_i} \, \| \, \rho_{\bar{a}} ] = - k \log p_i$, and so
\begin{equation}
S[\rho_{\bar{a}}] = \sum_{i = a}^n p_i \, S[\rho_{a_i}] - k \sum_{i = 1}^n p_i \log p_i.
\end{equation}

As outlined in the previous section, we are concerned with a system $A$ evolving between times $t_0$ and $t_1$ according to a time-varying Hamiltonian, and interacting successively with one or more  heat baths $\{B_i\}$, which initially have canonical distributions at temperatures $T_i$.  The Hamiltonians of the heat baths remain fixed throughout the evolution. We define
\begin{equation}
\ex{Q_i} = -\Delta\ex{H_{B_i}} = -\left( \ex{H_{B_i}}_{\rho_{B_i}(t_1)} - \ex{H_{B_i}}_{\rho_{B_i}(t_0)} \right).
\end{equation}
This is the expectation value of the heat energy obtained by $A$ from $B_i$.

Our first theorem relates the entropies as defined in the previous section to the Gibbs/von Neumann entropies. Though a simple one, it is of fundamental importance in the foundations of statistical mechanics, and deserves to be called the Fundamental Theorem of Statistical Mechanics.\footnote{This is not a new theorem.  The classical version of it is found in \citet[pp. 160--164]{GibbsBook}, and the quantum version, in \citet[\S 128--130]{Tolman}. Nonetheless,  it is not as well-known in the philosophical literature on statistical mechanics and thermodynamics as it should be.  \citet{MaroneyLandauer} refers to it as a \emph{generalized Landauer principle}.}
\begin{proposition}\label{Fund} If $\mc{M}$ is a class of manipulations of the sort outlined in Section \ref{setup}, then, for any states $a$, $b$,
\[
S_\mc{M}( a \rightarrow b) \leq S[\rho_b] - S[\rho_a].
\]
\end{proposition}
The following are immediate corollaries of this.
\begin{corollary}\label{Law2} \textbf{The second law of statistical thermodynamics.} For any state $a$,
\[
S_\mc{M}( a \rightarrow a) \leq 0.
\]
\end{corollary}
\begin{corollary}\label{GibbsEnt} If $a$ and $b$ can be connected reversibly---that is, if
\[
S_\mc{M}(a \rightarrow b \rightarrow a) = 0,
\]
then
\[
S_\mc{M}(a \rightarrow b) = S[\rho_b] - S[\rho_a].
\]
\end{corollary}
Thus, the Gibbs/von Neumann entropy is the state function whose existence is guaranteed by the second law plus reversibility.

Now, to the Landauer principle.  If a manipulation $M$ takes each of the states $\{a_i\}$ to the same state $b$, then it must also take any probabilistic mixture $\bar{a}$ of these states to $b$.  Let $\bar{a}$ be a mixture of the states $\{a_i\}$ with weights $\{p_i\}$. The expectation value of   heat exchanged when  $M$ is applied to this mixture is a weighted average  of exchanges that would occur in the states $\{a_i\}$, and so
\begin{equation}\label{0}
\sigma_M(\bar{a}\rightarrow b) = \sum_{i=1}^n p_i \: \sigma_M(a_i \rightarrow b).
\end{equation}
We must have, of course,
\begin{equation}\label{1}
\sigma_M(\bar{a}\rightarrow b) \leq S_\mc{M}( \bar{a} \rightarrow b).
\end{equation}
By the Fundamental Theorem,
\begin{equation}\label{2}
S_\mc{M}( \bar{a} \rightarrow b) \leq  S[\rho_b] - S[\rho_{\bar{a}}].
\end{equation}
From (\ref{mix}), the right-hand side of this is
\begin{equation}\label{3}
S[\rho_b] - S[\rho_{\bar{a}}] = \sum_{i=1}^n p_i \left( S[\rho_b] - S[\rho_{a_i}] \right) - \sum_{i=1}^n p_i \, S[\rho_{a_i} \, \| \, \rho_{\bar{a}}].
\end{equation}
Employing the Fundamental Theorem again,
\begin{equation}\label{4}
S[\rho_b] - S[\rho_{a_i}] \leq - S_\mc{M}(b \rightarrow a_i).
\end{equation}
Combining (\ref{0}), (\ref{1}), (\ref{2}), (\ref{3}), and (\ref{4}) gives us
\begin{equation}
\sum_{i=1}^n p_i \, \sigma_M(a_i \rightarrow b) \leq
\\  -\sum_{i=1}^n p_i \, S_\mc{M}(b \rightarrow a_i)  - \sum_{i=1}^n p_i\,  S[\rho_{a_i} \, \| \, \rho_{\bar{a}}].
\end{equation}
Rearranging, and recalling  the definition (\ref{dissdef}) of the dissipations, we get
\begin{equation}
\sum_{i=1}^n p_i \, \delta_M(a_i \rightarrow b)   \geq \sum_{i=1}^n p_i \, S[\rho_{a_i} \, \| \, \rho_{\bar{a}}].
\end{equation}
Thus, we have the result,
\begin{proposition}\label{LP} For any manipulation $M$ that takes each of $\{a_i\}$ to $b$, and any positive numbers $\{ p_i \}$ such that
\[
\sum_{i = 1}^n p_i = 1,
\]
we have
\[
\sum_{i=1}^n p_i \, \delta_M(a_i \rightarrow b)  \geq \sum_{i=1}^n p_i \, S[\rho_{a_i} \, \| \, \rho_{\bar{a}}],
\]
where $\bar{a}$ is a mixture of states $\{ a_i \}$ with weights $\{ p_i \}$.
\end{proposition}
This is our general version of Landauer's principle.  If we apply this to the case in which the states $\{a_i\}$ are distinguishable, we get the following corollary.
\begin{corollary}\label{LPSG}
For any manipulation $M$ that takes each of a distinguishable set $\{a_i\}$ to $b$, and any positive numbers $\{ p_i \}$ such that
\[
\sum_{i = 1}^n p_i = 1,
\]
we have
\[
\sum_{i=1}^n p_i \, \delta_M(a_i \rightarrow b) \geq -k \sum_{i=1}^n p_i \log p_i.
\]
\end{corollary}
As shown in the Appendix, this is equivalent to the following,
\begin{corollary}
For any manipulation $M$ that takes each of a distinguishable set $\{a_i\}$ to $b$,
\[
\sum_{i=1}^n e^{-\delta_M(a_i \rightarrow b)/k}  \leq 1.\]
\end{corollary}
This is the version stated in the previous section.

\section{Approximating reversibility}\label{reverse}  The second law of statistical  thermodynamics entails that, for any  $a$, $b$,
\begin{equation}
S_\mc{M}(a \rightarrow b \rightarrow a) \leq 0.
\end{equation}
We do not expect there to be  any process that takes $a$ to $b$ and then back to $a$ without any dissipation.  However, if the array of permitted manipulations is sufficiently rich, there maybe no bound on dissipation short of zero, and we may have $S_\mc{M}(a \rightarrow b \rightarrow a) = 0$.

One way to have a process that proceeds with negligibly small  dissipation is to keep the system $A$ in contact with a heat reservoir large and noisy enough that the reservoir  may be regarded as canonically distributed throughout the process, and to vary the parameters $\boldsymbol{\lambda}$ slowly enough that the time it takes for any appreciable change in these parameters is long compared to the equilibration time-scale of the system $A$.  Then the system $A$ may be treated as if it is  in equilibrium with the reservoir at each stage of the process.\footnote{This does not, of course, mean that it \emph{is} in equilibrium, only that, for the purposes at hand, differences between
quantities calculated on the basis of the equilibrium distribution and quantities calculated on the basis of the actual distribution are small enough that they may be neglected.}  We can also consider slowly varying the temperature of the reservoir.  For a process like that, at any time $t$ during  the process  $A$ may be treated as  having a canonical distribution for the instantaneous parameter values $( \boldsymbol{\lambda}(t), \beta(t))$.

If $\rho_1$ is a canonical distribution for parameters $( \boldsymbol{\lambda}, \beta )$, and $\rho_2$ a canonical distribution for slightly differing parameters  $( \boldsymbol{\lambda} + d \boldsymbol{\lambda}, \beta + d\beta )$, then, to first order in the parameter differences,\footnote{The classical version of this eq. (112) on p. 44 of \citet{GibbsBook}, and the quantum,  eq. (121.8) on p. 534 of  \citet{Tolman}. }
\begin{equation}
d \ex{H} = \ex{H_2}_{\rho_2} -  \ex{H_1}_{\rho_1} = \sum_i \left \langle \frac{\partial H}{\partial \lambda_i} \right \rangle_{\rho_1} d\lambda_i - \beta ^{-1} \, d \ex{\log{\rho}}.
\end{equation}
The first term on the right-hand side of this equation is  the expectation value of the work done in changing the external parameters;  the remainder is the expectation value of the heat obtained from the reservoir.
\begin{equation}\label{exQ1}
\ex{\dbar Q} = - k T \, d \ex{\log{\rho}},
\end{equation}
where $kT = \beta^{-1}$.  This means that, for a process in the course of which the system $A$ is in continual contact with a heat reservoir at temperature $T$ and the  parameters $\boldsymbol{\lambda}$ are varied slowly from values $\boldsymbol{\lambda}_a$ to $\boldsymbol{\lambda}_b$, the expectation value of total heat absorbed will have the approximate value
\begin{equation}\label{exQ2}
\ex{Q(a \rightarrow b)} \approx - k T ( \ex{\log{\rho_b}}_{\rho_{b}} - \ex{\log{\rho_a}}_{\rho_{a}}) = T \left(S[\rho_b] - S[\rho_a] \right).
\end{equation}
As long as there is no in-principle limit to how much time a state-transformation may take, there is no in-principle limit to how closely this approximation can hold, and equality will be approached as the time-scale of the changes in the parameters $\boldsymbol{\lambda}$ is increased, relative to the time-scale of equilibration of the system $A$.

The result (\ref{exQ2}) is a result about expectation values. It is \emph{not} assumed that the actual value of heat exchanged will be close to its expectation value, or even that it will \emph{probably} be close to its expectation value.  The probability distribution for the heat exchange may have a large variance, and probabilities of large deviations from the expectation value may be far from negligible. That is,  the   result does \emph{not} depend on disregard of fluctuations.  When we say that the system has time to equilibrate, this does not mean that it is ever in  a quiescent state, only that its distribution may be treated as canonical at each stage of the process.

Let $a$, $b$ be canonical states  with parameters  $( \boldsymbol{\lambda}_a, \beta_a )$, $(\boldsymbol{\lambda}_b, \beta_b )$.   We will say that a class of manipulations $\mc{M}$ connects $a$ and $b$ quasi-statically if
\begin{enumerate}
\item $\mc{M}$ contains manipulations of the following form
\begin{enumerate}
\item During time interval $[t_0, t_0 + T]$, the parameters undergo smooth evolution $\boldsymbol{\lambda}(t)$, with  $\boldsymbol{\lambda}(t_0) = \boldsymbol{\lambda}_a$ and $\boldsymbol{\lambda}(t_0 + T) = \boldsymbol{\lambda}_b$.
\item At time $t$ the system $A$ is in thermal contact with a heat reservoir at inverse temperate $\beta(t)$, where $\beta(t)$ is a smooth function with $\beta(t_0) = \beta_a$ and $\beta(t_0 + T) = \beta_b$.
\end{enumerate}
\item For any such manipulation, there is one that proceeds twice as slowly.  That is, there is a manipulation that takes place in time interval $[t_0, t_0 + 2T]$, with parameter values $\lambda'$, $\beta'$ where
\[
\boldsymbol{\lambda}'(t_0 + t) = \boldsymbol{\lambda}(t_0 + t/2); \quad \beta'(t_0 + t) = \beta(t_0 + t/2).
\]
for $t \in [0, 2T]$.
\end{enumerate}

Then we have the following result.
\begin{proposition}\label{quasi} If $a$, $b$ are canonical states, and $\mc{M}$ is a class of manipulations that connects $a$ to $b$ quasi-statically, then
\[
S_\mc{M}(a \rightarrow b) =  S[\rho_b] - S[\rho_a].
\]
\end{proposition}
We have, as a trivial corollary,
\begin{corollary}
If $a$, $b$ are canonical states, and $\mc{M}$ is a class of manipulations that connects $a$ to $b$ quasi-statically, and also connects $b$ to $a$ quasi-statically, then
\[
S_\mc{M}(a \rightarrow b \rightarrow a) =  0.
\]
\end{corollary}

Suppose that we have a system to which can be applied a manipulable external potential $V_{\boldsymbol{\lambda}}$, and which can also be confined, by suitable barriers, to various regions $\{\Gamma_i\}$ of its state space.  Let $\{a_i\}$ be a finite set of canonical states, confined to the regions $\{\Gamma_i\}$, with values $\boldsymbol{\lambda}_a$ of the manipulable parameters $\boldsymbol{\lambda}$ on which the external potential depends, and let  $\{b_i\}$ be a set of canonical distributions  confined to the same regions, with parameter  values $\boldsymbol{\lambda}_b$.  Then, for any desired degree of approximation to the quasistatic limit, we can find a sufficiently slow variation of the parameters $\boldsymbol{\lambda}$ that yields the desired degree of approximation for \emph{all} of the transitions $a_i \rightarrow b_i$.  We will say, of such a situation, that $\mc{M}$ uniformly quasi-statically connects  $\{a_i\}$ to  $\{b_i\}$.  We have, as another corollary to Proposition (\ref{quasi}).
\begin{corollary}
Let $\{a_i\}$, $\{b_i\}$ be sets of canonical states, such that  $\mc{M}$ uniformly quasi-statically connects $\{a_i\}$ to  $\{b_i\}$ and $\{b_i\}$ to  $\{a_i\}$. Let  $\{p_i\}$ be a set of non-negative numbers that sum to 1, and let  $\bar{a}$ and $\bar{b}$ be probabilistic mixtures of  $\{a_i\}$ and $\{b_i\}$ with weights $\{p_i\}$.  Then
\[
S_\mc{M}(\bar{a} \rightarrow \bar{b} \rightarrow \bar{a}) =  0.
\]
\end{corollary}

\section{Example: One-particle gas}
 The simplest example I can think of for illustrating erasure that is  a single particle in a box, with a partition that can be inserted and removed.  If this is the only available manipulation, $S_\mc{M}(a \rightarrow b)$ will be zero for all states $a$, $b$ of the same temperature.  To get nontrivial entropies, we need to introduce the possibility of doing work on and obtaining work from the system.

Suppose that the particle can be subjected to an external potential $V_\lambda$, that varies in the $x$-direction only.  We take the system to be in thermal equilibrium with a heat bath at temperature $T$.  On a canonical distribution, the  distributions of the momentum $\mb{p}$ and the coordinates other than $x$ are unchanged when the potential $V_\lambda$ is varied.  We  therefore integrate these out, and consider the marginal distribution of the coordinate $x$.
\begin{equation}\label{particledistribution}
\rho_{\lambda,\beta}(x) =
\left\{
\begin{array}{lll}
Z_{\lambda, \beta}^{-1} \, e^{-\beta \, V_\lambda(x)},& \quad &\mbox{inside the container;}
\\
0. & \quad & \mbox{outside}.
\end{array}
\right.
\end{equation}
Take the   $x$-coordinate within the container to range from $-l$ to $l$. The partition function is
\begin{equation}
Z_{\lambda, \beta}  = \int_{-l}^{l} e^{-\beta \, V_\lambda(x)} \, dx.
\end{equation}
We need not assume that the potential $V_\lambda$ is under perfect control.  It, too, may fluctuate, with its own probability distribution. Evolution of a probability distribution, via the Liouville equation, of a system subject to a  potential $V$ that fluctuates with a probability distribution of its own, independent of the state of the system,  is the same as  evolution under a steady potential equal to the expectation value $\ex{V}$ of the potential.  Thus, if the external force fluctuates,  the stable distribution is the same as (\ref{particledistribution}), with $V_\lambda(x)$ replaced by its expectation value at the point $x$.  Fluctuations of the external potential, even large ones, do not invalidate our analysis.

Suppose the force on the particle is constant within the box, and may be varied in both strength and direction.  The particle could, for example, be a charged particle, and the applied field an electric field.  Then the external potential varies linearly with $x$.  Take it to be,
\begin{equation}
V_\lambda(x) = \lambda  k T \, x/l.
\end{equation}
where $\lambda$ is a dimensionless parameter.

The analog of compressing or expanding the one-particle gas is varying the external potential.  As $\lambda$ is increased from zero, the distribution of the particle becomes more and more concentrated towards the left end of the container.  We can make the probability that it is to the left of any chosen location as high as we want by taking $\lambda$ sufficiently large.  Similarly, for negative values of $\lambda$, the distribution is concentrated towards the right end of the container.

Relative to a canonical distribution with $\lambda = 0$, a distribution for a large value of $\lambda$ has  a large value of free energy, and so we have to do work on the gas while increasing the potential.   The work done may be recovered by decreasing the potential back to zero.  If the process is done slowly enough that the particle can be treated as canonically distributed at each state of the process, the expectation value of the work recovered while decreasing the potential is equal to the expectation value work of the work  done in increasing it.

Let $b$ be a state in which no partition is present and the applied potential is zero. The probability distribution of the particle is evenly distributed throughout the container.   Now insert a partition that divides the container into subvolumes with ratio $p:(1 - p)$.  Let  $a_1(p)$ be a state in which the particle is to the left of the partition, and let $a_2(p)$ be a state in which the particle is to the right of the partition.

The states $a_1(p)$ and  $a_2(p)$ are perfectly distinguishable states.  There's a complication, however: given our class of manipulations, we have no way to prepare them, starting from state $b$.  If we start from $b$ and increase the potential, we can make the probability that the particle is to the left of where we intend to drop the partition as high as we like, but it can never be equal to 1.

In place of these states  $a_1(p)$ and  $a_2(p)$,  which are perfectly distinguishable but not preparable using the manipulations considered, we  consider a pair of states that are \emph{almost} distinguishable, and are preparable. Let $\epsilon$ be a small positive number, and let $a^\epsilon_1(p)$ be a state in which $V_\lambda$ is zero, and a partition is present, dividing the container into subvolumes with ratio $p:(1-p)$, and in which there is a probability of $1 - \epsilon$ that the particle is to the left of the partition, and probability $\epsilon$ that it is to the left. Define $a^\epsilon_2(p)$ similarly, with the probabilities reversed.

One manipulation that takes $a^\epsilon_1(p)$ to  $b$ is removal of the partition, after which the particle equilibrates.  This is inefficient, as we could have  performed an expansion of the gas, in the course of which work is obtained and heat enters the gas from the reservoir.

To see how much inefficiency, we consider the following process, which is analogous to a controlled expansion of a gas.  We start in state $a^\epsilon_1(p)$.
\begin{enumerate}
\item We first slowly increase $\lambda$ to the point at which, on the canonical distribution for $V_\lambda$, the particle has probability $1 - \epsilon$ of being to the left of the partition, and probability $\epsilon$ of being on the right.
\item We remove the partition, allowing the particle to move freely throughout the container. The probability distribution does not change, as the probability, on the equilibrium distribution,  of the particle being on the left of the former location of the partition is the same as it was before the partition was removed.\footnote{General rule: if we take  state space $\Gamma$ and   partition the  space into disjoint regions $\Gamma_i$, a canonical distribution $\rho$ defined on $\Gamma$ is a mixture of canonical distributions $\rho_i$  confined to the regions $\Gamma_i$, with weights  being the probabilities, on $\rho$, that the system is in $\Gamma_i$.}
\item The potential is slowly decreased to zero.
\end{enumerate}
The process can be performed in reverse order to create $a^\epsilon_1(p)$ from $b$.  If we have available to us arbitrarily slow processes,
\begin{equation}
S_\mc{M}(a^\epsilon_1(p) \rightarrow b \rightarrow a^\epsilon_1(p)) = S_\mc{M}(a^\epsilon_2(p) \rightarrow b \rightarrow a^\epsilon_2(p)) = 0.
\end{equation}
The expectation value of heat gained in the  process of expansion is, in the quasistatic approximation,
\begin{multline}
\ex{Q(a^\epsilon_1(p) \rightarrow b)} = T ( S[b]  -  S[a^\epsilon_1(p)])
\\ = -kT [(1 - \epsilon ) \log p + \epsilon \log(1-p) - v(\epsilon)],
\end{multline}
where
\begin{equation}
v(\epsilon) = \epsilon \log \epsilon + (1 -\epsilon) \log(1- \epsilon).
\end{equation}
We can make $\ex{Q(a^\epsilon_1(p) \rightarrow b)}$ as close to $-kT \log p$ as we like by taking $\epsilon$ sufficiently small.

Therefore, erasure by removing the partitions has associated with it inefficencies,
\begin{equation}\label{dissipations}
\begin{array}{l}
\eta_1  = -k[(1 - \epsilon ) \log p + \epsilon \log(1-p) - v(\epsilon)] \approx -k \log p,
\\ \\
\eta_2 = -k[\epsilon \log p + (1 - \epsilon) \log(1-p) - v(\epsilon)] \approx -k \log(1-p).
\end{array}
\end{equation}

Suppose that we want an erasure process that takes both $a^\epsilon_1(p)$ and $a^\epsilon_2(p)$ to the state $b$.  One such process goes by removal of the partition.  This has the inefficiencies exhibited in (\ref{dissipations}).   But we have only availed ourselves of a fairly limited set of operations. Would it be possible to concoct a different set of operations, which might include the employment of auxiliary systems subject to any sort of Hamiltonian we might dream up, whether or not realization of such Hamiltonians is remotely feasible, and thereby construct an operation that takes both $a^\epsilon_1(p)$ and $a^\epsilon_2(p)$ to $b$,   with lower  inefficiency for both input states than the lossy removal-of-partition operation?  Alas, the answer is negative. As the reader can verify, as long as $\epsilon < p < 1 - \epsilon$, the pair of inefficiencies (\ref{dissipations})  saturate the Landauer bound exhibited in Proposition \ref{LP}.  This means that no process, no matter how elaborate, will achieve a lower inefficiency for both input states,   so long as all exchanges of heat are with canonically distributed reservoirs, there are at the beginning of the process no dynamically relevant correlations between the state of $A$ and  either the auxiliary systems or the reservoirs,  the evolution of the total system is Hamiltonian, and at  the end of the evolution the auxiliary systems are restored to their initial states.

\section{The LPSG proof vindicated}  The LPSG proof proceeds as follows.  Suppose we have a manipulation $M_L$ that takes each of a distinguishable set of states $\{a_i, i = 1,\ldots, n\}$ of a device $D$ to a common destination state $b$.  The proof employs as an auxiliary system a one-molecule gas in a box into which partitions may be inserted and removed, and which can be expanded reversibly. LPSG  reason that, on pain of violating the statistical second law of thermodynamics, the manipulation $M_L$ must satisfy the Landauer principle. This involves considering the following cycle of operations (performed with both the device $D$ and the gas $G$ in contact with a heat reservoir at all times).  The starting state is one in which device $D$ is in state $b$, and there are no partitions in the box.
\begin{enumerate}
\item{$n-1$ partitions are inserted into the box, dividing its volume into $n$ subvolumes, with volumes that are fractions $p_i$ of the total volume.  With probability $p_i$, the gas molecule is in the $i$th subvolume.}
\item{A controlled operation is performed on $D$, using the state of the gas $G$ as control.  If the gas molecule is in the $i$th subvolume, $b$ is taken to $a_i$.}
\item{A controlled operation is performed on the gas $G$, using the state of $D$ as control.  The $ith$ subvolume is expanded reversibly, obtaining heat
 $-k T \log p_i$ from the reservoir. The gas has now been restored to its initial state.}\label{Expand}
\item{The operation $M_L$ is performed, restoring the device $D$ to the state $b$.}
\end{enumerate}
If one works through the expectation values of heat exchanges in the course of this cycle, assuming the statistical second law but not assuming reversibility of  the processes $b \rightarrow a_i$, then what is obtained is precisely our Corollary \ref{LPSG} of section \ref{proving}.  Obviously, if one replaces the assumption that heat $-kT \log p_i$ can be obtained in step \ref{Expand} with the assumption that there are operations such that the expectation value of heat obtained can come arbitrarily close to  $-kT \log p_i$,  the result still obtains.

The point of contention is whether expansion of a one-molecule gas can be performed in such a way that expectation value of heat obtained is arbitrarily close to $-kT \log p_i$.  Norton, in the works cited, contends that this is false.  In my opinion \citet{LRCircles} are right when they say that he  has not established this.  However, if one has doubts about this being true for a one-molecule gas  expanded by a piston, because of lack of control over a sufficiently sensitive piston, our example from the previous section of a one-molecule gas subjected to an external potential  may be substituted.

We replace step \ref{Expand} with the following process.  For simplicity we illustrate it for the case of a single partition; extension to multiple partitions is straightforward.  Suppose the particle is found to be to the left of the partition. The initial state is $a_1(p)$.
\begin{enumerate}
\item Slowly increase $\lambda$ to a high positive value $\lambda^*$.
\item Remove the partition, and allow the system to equilibrate.  Some heat is absorbed from the reservoir, but, for large $\lambda^*$, this is small.
\item Slowly decrease $\lambda$ to zero.
\end{enumerate}
If the particle is found to the right of the partition, one takes $\lambda$ to a large negative value instead.  It is not difficult to calculate the expectation value of heat obtained in such a  process in the adiabatic limit.  The details of this calculation need not concern us; what matters if that it can be made arbitrarily close to $-kT \log p$ by taking $\lambda^*$ sufficiently large.\footnote{For those who are interested, the result is
\[
\ex{Q} = - kT \log p - kT \log \left(\frac{1 - e^{-2\lambda^*}}{1 - e^{-2 p \lambda^*}} \right).
\]
For any $p$, $0 < p < 1$, for large $\lambda^*$ we have
\[
\ex{Q} \approx   - kT \log p  - kT  e^{-2 p \lambda^*}.
\]
Therefore, $\ex{Q}$ approaches $- kT \log p$ exponentially with increase of $\lambda^*$.
}

\section{Conclusion}  Landauer's principle is a theorem of statistical mechanics.  The worries raised by Norton about assuming reversibility can be addressed; fluctuations pose no threat to the extent we can approximate reversibility, in the relevant sense.  If the system being manipulated is in contact   with a heat reservoir at temperature $T$ throughout a cycle of operations, the \emph{expectation value} of heat exchanged over the course of the cycle can be made as small as one likes if one is patient enough.  On any given run of the cycle, the actual heat exchanged may differ wildly from this expectation value, but it is the expectation value that is relevant to the statistical version of Landauer's principle.

\section{Acknowledgements}  I am grateful to a number of people with whom I have discussed these matters over the years.   In particular, I thank  Owen Maroney for drawing my attention to what I have called the Fundamental Theorem,  John Norton for discussions of reversible processes, and Katie Robertson for comments on an earlier draft of this article.

\section{Appendix}
\subsection{Proof of the Fundamental Theorem} To be proven: If $\mc{M}$ is a class of manipulations of the sort outlined in Section \ref{setup}, then, for any states $a$, $b$,
\[
S_\mc{M}( a \rightarrow b) \leq S[\rho_b] - S[\rho_a].
\]
We use the following lemmas.
\begin{lemma}\label{FreeMin}
For any Hamiltonian $H$, and any $T > 0$, the canonical distribution at temperature $T$ minimizes
\[
\ex{H}_\rho - T S[\rho].
\]
\end{lemma}
\begin{lemma}\label{subadd} \emph{Subadditivity.}  For a composite system $AB$,
\[
S[\rho_{AB}] \leq S[\rho_A] + S[\rho_B],
\]
with equality if and only if the subsystems are probabilistically independent.
\end{lemma}
\begin{lemma}\label{Scons}
$S[\rho]$ is conserved under Hamiltonian evolution.
\end{lemma}
 We consider some manipulation $M \in \mc{M}$  that takes a state $a$ of $A$ at $t_0$ to a state $b$ at $t_1$.  At time $t_0$ the composite system consisting of $A$ and $\{B_i\}$ has distribution represented by density $\rho_{tot}(t_0)$.  At time $t_1$ the density is    $\rho_{tot}(t_1)$.  We will write $S_{tot}(t)$ as an abbreviation for $S[\rho_{tot}(t)]$, and similarly for $S_A(t)$ and $S_{B_i}(t)$.

   By Lemma \ref{FreeMin} we have, for each reservoir $B_i$,
\begin{equation}
\ex{H_{B_i}(t_0)} - T_i S_{B_i}(t_0) \leq \ex{H_{B_i}(t_1)} - T_i S_{B_i}(t_1),
\end{equation}
or,
\begin{equation}
\Delta \ex{H_{B_i}} - T_i \Delta S_{B_i} \geq 0.
\end{equation}
Since $\ex{Q_i} = -  \Delta \ex{H_{B_i}}$, this gives
\begin{equation}\label{QB}
\frac{\ex{Q_i}}{T_i} \leq -\Delta S_{B_i}.
\end{equation}
Because $A$ is uncorrelated with each $B_i$ at $t_0$,
\begin{equation}\label{f1}
S_{tot}(t_0) = S_A(t_0) + \sum_{i=1}^n  S_{B_i}(t_0).
\end{equation}
Because of subadditivity,
\begin{equation}\label{f2}
S_{tot}(t_1) \leq  S_A(t_1) + \sum_{i=1}^n  S_{B_i}(t_1).
\end{equation}
Because Hamiltonian evolution conserves $S$,
\begin{equation}\label{f3}
S_{tot}(t_1) = S_{tot}(t_0).
\end{equation}
Taken together, (\ref{f1}), (\ref{f2}), and (\ref{f3}) yield,
\begin{equation}
\Delta S_A + \sum_{i=1}^n \Delta S_{B_i} \geq 0.
\end{equation}
This, together with (\ref{QB}), gives us the result,
\begin{equation}
\sigma_M(a \rightarrow b)  =  \sum_{i=1}^n \frac{\ex{Q_i}}{T_i} \leq \Delta S_A.
\end{equation}
Since this must hold for every manipulation in the set $\mc{M}$, it must hold also for $S_\mc{M}(a \rightarrow b)$, which we defined as the least upper bound of the set of all $\sigma_M(a \rightarrow b)$ for $M \in \mc{M}$. This gives us the desired result,\
\begin{equation}
S_\mc{M}(a \rightarrow b) \leq \Delta S_A.
\end{equation}
\subsection{Proof of equivalence of two formulations.}
\begin{lemma}\label{eq} Let $\{x_i, i = 1,\ldots, n\}$ be any sequence of $n$ real numbers.  The following are equivalent.
\begin{enumerate}[label = \Alph*)]
\item For all non-negative  $\{p_i, i = 1,\ldots, n\}$ such that $\sum_i p_i = 1$,
\[
\sum_{i=1}^n p_i \, x_i \geq  -\sum_{i=1}^n p_i \log p_i.
\]
\item \[ \sum_{i=1}^n e^{-x_i} \leq 1.\]
\end{enumerate}
\end{lemma}

To prove this, we use the following.
\begin{lemma}\label{pq}
For any positive numbers $\{ p_i \}$, $\{ q_i \}$,
\[
\sum_{i=1}^n p_i \log q_i  - \log\left(\sum_i q_i\right) \leq \sum_{i =1}^n p_i \log p_i  - \log\left(\sum_i p_i \right).
\]
\end{lemma}
To prove this: given $\{p_i\}$, find $\{q_i\}$ that maximizes the LHS; this maximum value is the RHS.  Details omitted. We now proceed to the proof of  Lemma \ref{eq}.

\bigskip

\noindent \emph{Proof that $(A) \Rightarrow (B)$}.  Suppose that $\{ x_i \}$ are such that (A) holds.  Take
\begin{equation}
p_i = \frac{e^{-x_i}}{\sum_{j=1}^n e^{-x_j}}.
\end{equation}
Then $\sum_i p_i = 1$, and
\begin{equation}
\sum_{i=1}^n p_i \, x_i =  -\sum_{i=1}^n p_i \log p_i  -  \log \left(\sum_{j=1}^n e^{-x_j} \right).
\end{equation}
In order for (A) to be satisfied, we must have
\begin{equation}
\log \left(\sum_{j=1}^n e^{-x_j} \right) \leq 0,
\end{equation}
which is equivalent to
\begin{equation}
\sum_{j=1}^n e^{-x_j} \leq 1.
\end{equation}

\noindent \emph{Proof that $(B) \Rightarrow (A)$}.  Suppose that $\{ x_i \}$ are such that (B) holds. Let $q_i = e^{-x_i}$.  Then
\begin{equation}
\sum_{i=1}^n p_i x_i = - \sum_{i=1}^n  p_i \log q_i.
\end{equation}
By Lemma \ref{pq}, for any $\{ p_i \}$ such that $\sum_i p_i = 1$,
\begin{equation}
- \sum_{i=1}^n p_i \log q_i \geq - \sum_{i=1}^n p_i \log p_i - \log \left( \sum_{i=1}^n q_i  \right),
\end{equation}
and so
\begin{equation}
\sum_{i=1}^n p_i x_i \geq - \sum_{i=1}^n p_i \log p_i - \log \left( \sum_{i=1}^n q_i  \right).
\end{equation}
Because of (B),
\begin{equation}
\log \left( \sum_{i=1}^n q_i  \right) = \log \left( \sum_{i=1}^n e^{-x_i} \right) \leq 0,
\end{equation}
and so,
\begin{equation}
\sum_{i=1}^n p_i x_i \geq - \sum_{i=1}^n p_i \log p_i.
\end{equation}

\newpage
\bibliographystyle{chicago}

\end{document}